\documentclass[aps,prl,twocolumn,superscriptaddress]{revtex4-1}

\usepackage{graphicx}

\newcommand{\mr}[1]{\mathrm{#1}}

\def\<{\left\langle} \def\>{\right\rangle} \def\({\left(} \def\){\right)}
\def\Gi{G_\infty}
\def\et{\textit{et al. }}
\def\inv{$^{-1}$}

\begin{document}


\title{Measurement of angular momentum transport in turbulent flow between independently rotating cylinders}


\author{M. S. Paoletti}
\affiliation{Departments of Physics, Geology, Institute for Research in Electronics
and Applied Physics, and}
\author{D. P. Lathrop}
\email[email address: ]{lathrop@umd.edu}
\affiliation{Departments of Physics, Geology, Institute for Research in Electronics
and Applied Physics, and}
\affiliation{Institute for Physical Sciences and Technology, University of Maryland, College Park, MD 20742}


\begin{abstract}
We present measurements of the angular momentum flux (torque) in Taylor-Couette flow of water between independently rotating cylinders for all regions of the $\(\Omega_1, \Omega_2\)$ parameter space at high Reynolds numbers, where $\Omega_1$ $\(\Omega_2\)$ is the inner (outer) cylinder angular velocity.  We find that the Rossby number $Ro = \(\Omega_1 - \Omega_2\)/\Omega_2$ fully determines the state and torque $G$ as compared to $G(Ro = \infty) \equiv \Gi$.  The ratio $G/\Gi$ is a linear function of $Ro^{-1}$ in four sections of the parameter space.  For flows with radially-increasing angular momentum, our measured torques greatly exceed those of previous experiments [Ji \textit{et al.}, Nature, \textbf{444}, 343 (2006)], but agree with the analysis of Richard and Zahn [Astron. Astrophys., \textbf{347}, 734 (1999)].
\end{abstract}


\maketitle

Rapidly-rotating shear flows are ubiquitous in geophysical and astrophysical settings such as planetary atmospheres, stellar interiors and accretion disks.  In order for essential fundamental processes to occur, like matter inflow towards compact objects \cite{shakura73}, there must be an exchange of angular momentum through such flows.  Determining the flux of angular momentum in rotating shear flows is difficult and has been actively studied \cite{lathrop92,lewis99,richard99,hure01,richard03,dubrelle05,ji06,eckhardt07a,ravelet10} since the initial measurements of Wendt \cite{wendt33} and Taylor \cite{taylor36}.  While shear tends to destabilize fluid flows, rotation stabilizes them whenever angular momentum increases radially outward ($dL/dr > 0$, the Rayleigh criterion \cite{rayleigh16}); although, shear turbulence has been shown to occur even in Rayleigh-stable systems \cite{coles65}.   Most astrophysical flows are Rayleigh-stable, which in the absence of other instabilities cannot transport angular momentum efficiently enough to allow for the breadth of observed phenomena.

The question remains as to how astrophysical objects transport angular momentum at the observed rates \cite{shakura73}.  It is widely accepted that turbulence is mediated by an instability, which enhances angular momentum transport; although, the particular driving mechanism remains controversial.  In the case of electrically-conducting accretion disks, magnetic fields can give rise to the linear magneto-rotational instability (MRI) \cite{velikhov59,chandrasekhar60,balbus91,sisan04}.  The MRI increases the flux of angular momentum, but it is still unknown if (\textit{i}) it is the only instability and (\textit{ii}) it is sufficient to produce the observed angular momentum transport.

Ji \textit{et al.} \cite{ji06} measured the flux of angular momentum in Taylor-Couette flow between two cylinders that rotated with angular velocities $\Omega_1$ and $\Omega_2$ for the inner and outer boundaries, respectively.  They focussed on \lq\lq quasi-Keplerian" flows in water, where the angular velocity decreases radially $\(d\Omega/dr < 0\)$ while the angular momentum increases $\(dL/dr > 0\)$, as in many astrophysical flows.  Their measurements of the flux of angular momentum, inferred from velocity measurements, were indistinguishable from those in solid-body rotation ($\Omega_1 = \Omega_2$), where the flux is zero.  Ji \textit{et al.} concluded that there was no hydrodynamic instability and that non-magnetic, quasi-Keplerian flows at Reynolds numbers up to $2\times 10^6$ are \lq \lq essentially steady."  These results are at odds with the analysis of Dubrelle \et \cite{dubrelle05}, who used prior velocity measurements to infer that the flux of angular momentum is nonzero for Rayleigh-stable flows.  However, there has yet to be a study that directly measures the flux of angular momentum in all portions of the $\(\Omega_1,\Omega_2\)$ parameter space.  Furthermore, Ji \et could not quantify the flux for quasi-Keplerian flows since the random errors in their measurements exceeded the measured values.

In this Letter, we characterize the flux of angular momentum (torque) in Taylor-Couette flow between independently rotating cylinders for all regions of the parameter space up to Reynolds numbers $Re = \(\Omega_1 - \Omega_2\)(b-a)(a+b)/2\nu = 4.4 \times 10^6$, where $\nu$ is the kinematic viscosity and $a$ ($b$) is the inner (outer) radius.  Rather than inferring the torque from velocity measurements, as in \cite{ji06}, we directly measure the torque required to drive the inner cylinder \cite{lathrop92}.  The precision of our measurements allows us to distinguish between solid-body rotation and other flow states observed, in contrast to ref.\ \cite{ji06}.

\begin{figure}
\begin{center}
\includegraphics[width=7.5cm]{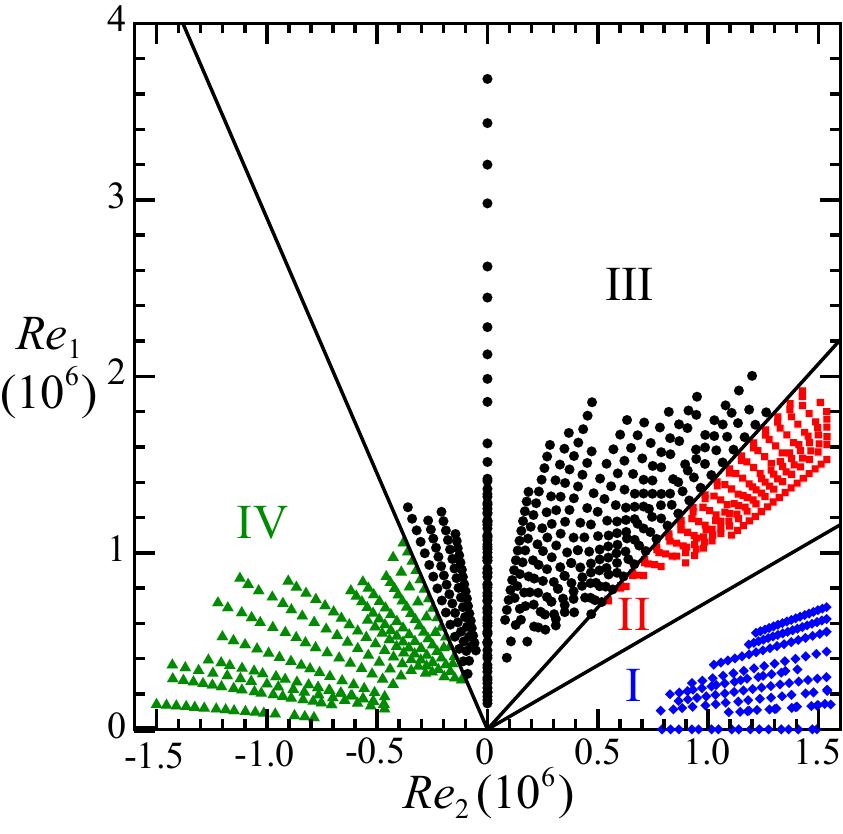}%
\caption{Our experiments span a large range of the $\(Re_2,Re_1\)$ parameter space, which we divide into regions.  Region I (blue diamonds) is defined as $-1 \leq Ro < 0$.  Region II (red squares) has $0 < Ro < \eta^{-2} -1$, where $Ro = \eta^{-2}-1$ defines the Rayleigh stability criterion \cite{rayleigh16}.  Region III (black circles) is for $Ro < -4$ and $\eta^{-2}-1 < Ro$.  Finally, region IV (green triangles) has $-4 < Ro < -1$.  Data are not acquired very near $Ro = 0$, since the torques are comparable to our measurement precision of 0.01~Nm ($G \sim 10^8$ at 50~$^{\circ}$C).}
\label{param_space}
\end{center}
\end{figure}

We partition the experimental parameter space (Fig.\ \ref{param_space}) into four regions.  The regions are defined by the Rossby number, which we claim is the main controlling parameter for the dynamics:
\begin{equation}
Ro \equiv \(\Omega_1 - \Omega_2\)/\Omega_2 = \(Re_1/\eta Re_2\)-1,
\label{rossby_eq}
\end{equation}
where $\eta = a/b$, $Re_1 = \Omega_1 a (b-a)/\nu$ and $Re_2 = \Omega_2 b (b-a)/\nu$ are the radius ratio and inner and outer cylinder Reynolds numbers. The boundaries co-rotate in region I with $0 \leq \Omega_1 < \Omega_2$.  Region II also has co-rotation with radially increasing (decreasing) angular momentum $L$ (angular velocity $\Omega$), as in quasi-Keplerian flows.  Regions II and III are divided by the Rayleigh stability criterion $dL/dr = 0$.  Region IV is characterized by counter-rotation with $-4 < Ro < -1$.  Regions I and II are Rayleigh-stable and are separated at $Ro = 0$ (solid-body rotation).  Regions III and IV are linearly unstable since $dL/dr < 0$.  The division between regions~III and IV is a result of the observation of a maximum in the measured torque at $Ro = -4$.  The dynamics are symmetric under negating both $\Omega_1$ and $\Omega_2$; therefore, regions I--IV describe the entire Taylor-Couette parameter space.

Our experiments use water as the working fluid and are conducted in the apparatus constructed by Lathrop \textit{et al.} \cite{lathrop92}, which has been modified to allow the outer cylinder to rotate independently.  The acrylic outer cylinder used in previous experiments \cite{lathrop92,lewis99} has been replaced by an anodized aluminum cylinder with the same inner radius $b = 22.085$~cm and length $L = 69.50$~cm.  The inner cylinder is stainless steel with a radius $a = 16.000$~cm yielding a radius ratio $\eta = a/b = 0.7245$ and an aspect ratio $\Gamma = L/(b-a) = 11.47$.  The inner cylinder is rotated up to $\Omega_1/2\pi = 20$~s$^{-1}$ while the outer cylinder may be rotated in either direction up to $\left|\Omega_2/2\pi\right| = 10$~s$^{-1}$.  Both angular velocities are measured precisely by shaft encoders and controlled to within 0.2\% of the set value.

The axial boundaries rotate with the outer cylinder, although the ideal Couette geometry is unbounded axially.  To avoid end effects in our torque measurements, the inner cylinder is divided axially into three sections of length 15.69, 40.64 and 15.69~cm (see Fig.\ 3 of \cite{lathrop92}).  Only the central section of the inner cylinder senses the torque of the fluid as described in ref.\ \cite{lathrop92}.  Therefore $2.58(b-a)$ from each of the axial boundaries, where secondary circulation setup by finite boundaries (Ekman pumping) is strongest, are avoided in the torque measurements.

The local wall shear stress is measured at the outer boundary using a flush-mounted hot film probe.  The probe is located at the mid-height of the experiment.  The measurements are performed in the constant temperature mode using a Dantec mini-CTA anemometer.  The frequency response of the probe exceeds our sampling rate of 10~kHz.  The shear stress measurements are calibrated \textit{in situ} using the method described in ref.\ \cite{lathrop92}.

The desired accuracy of our measurements requires that the temperature of the water be precisely controlled.  In contrast to prior experiments \cite{lathrop92,lewis99} where the system was cooled at the axial boundaries, we control the temperature through the outer cylinder.  This procedure is superior owing to the 6.5~fold increase in temperature-controlled surface area.  Furthermore, the working fluid is now temperature-controlled along the entire axial length of the experiment.  This is particularly important for the flows in regions I and II of the parameter space, where mixing is greatly reduced.  In all of our measurements the temperature is controlled to within 0.02~$^{\circ}$C of 50~$^{\circ}$C yielding a kinematic fluid viscosity of $\nu = 0.0055$~cm$^2$/s, except for $Re > 2 \times 10^6$ where $T = 90$~$^{\circ}$C and $\nu = 0.0032$~cm$^2$/s.  This control algorithm and temperature range would not be possible with an acrylic outer cylinder owing to the poor thermal properties compared to those of aluminum.

\begin{figure}
\begin{center}
\includegraphics[width=8.6cm]{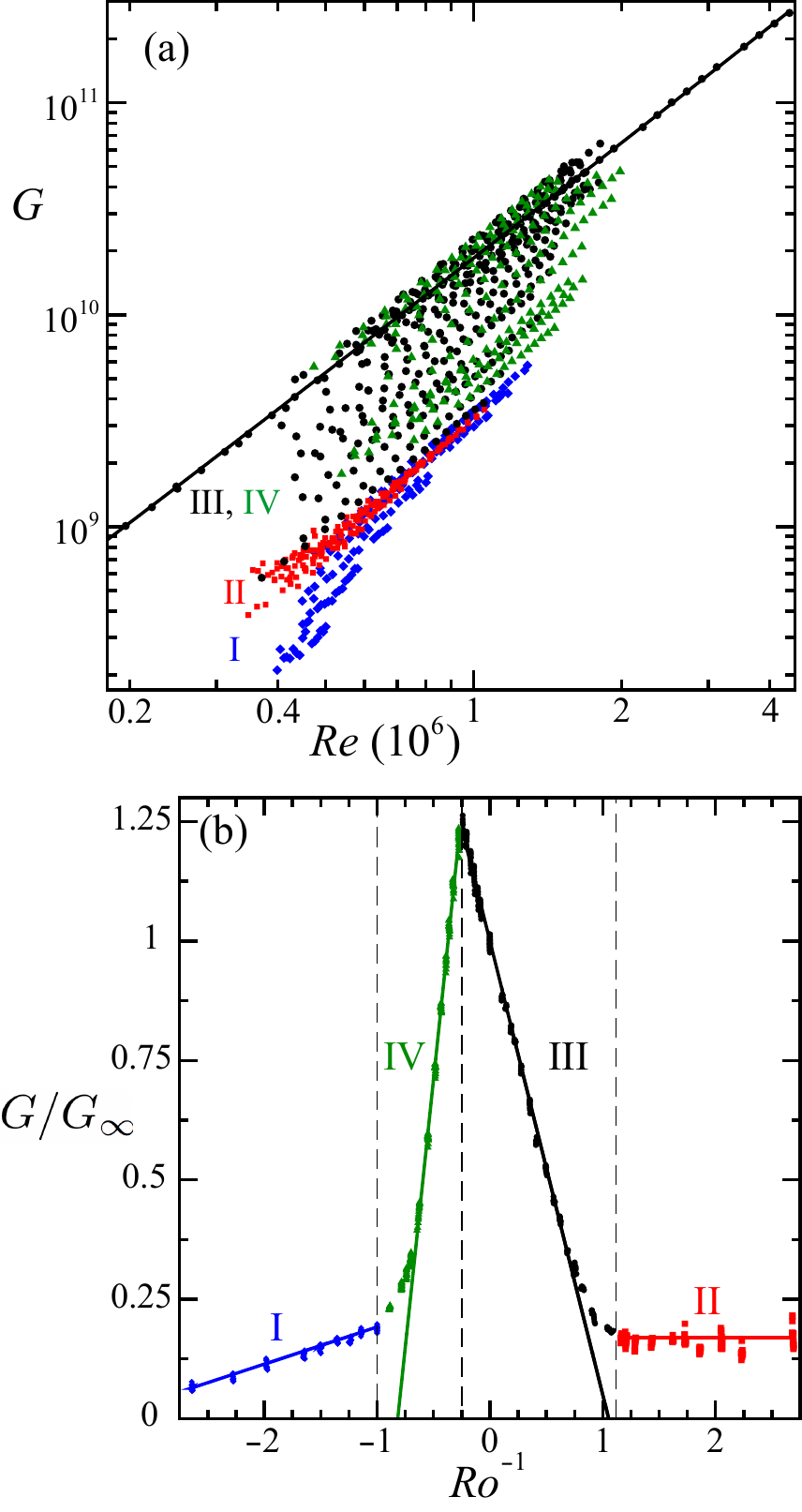}%
\caption{Experimental measurements of the dimensionless torque $G$ as a function of (a) $Re$ and (b) $Ro^{-1}$ with the symbols defined in Fig.\ \ref{param_space}.  The solid line in (a) represents the fit $\Gi$ for $Ro = \infty$ given in eq.\ (\ref{Ginf_eq}) \cite{lewis99}.  The solid lines in (b) correspond to the fits given in eqs. (\ref{Gfit1})--(\ref{Gfit4}).}
\label{Gfig}
\end{center}
\end{figure}

We study the scaling of the torque as a function of $Re$ for several values of $Ro$.  The measured torque $\tau$ is made dimensionless by defining $G = \tau/\rho \nu^2 L_{\mr{c}}$, where $\rho$ is the fluid density and $L_{\mr{c}}$ the length of the torque-sensing central section of the inner cylinder.  Our measured values of $G$ are shown in Fig.\ \ref{Gfig}(a).  We note that both $Re$ and $G$ are negative in region I and we therefore plot their absolute values.  We compare our data to the best fit of $G(Ro=\infty)$ (solid curve) given in ref.\ \cite{lewis99}, which is well described by the following:
\begin{equation}
\frac{Re_1}{\sqrt{\Gi}} = 1.56 \log \sqrt{\Gi} - 1.83.
\label{Ginf_eq}
\end{equation}
Our measurements of $G$ obey the scaling observed in previous experiments for $Ro = \infty$ \cite{lathrop92,lewis99}, even above the previous maximum of $Re=1.2\times 10^6$.  However, depending upon $Ro$, the value of $G$ for a given $Re$ may be higher or lower than $\Gi$ \cite{dubrelle05,ravelet10}.  The main observed dependence on $Ro$ is a vertical shift in this representation.

To determine this shift in $G$ with $Ro$ we measure $G/G_{\infty}$, as in ref. \cite{dubrelle05}, as a function of $Ro^{-1}$, where $G_{\infty}$ is given in eq.~(\ref{Ginf_eq}) \cite{lewis99}.  Figure \ref{Gfig}(b) shows that $G/G_{\infty}$ is essentially constant for each value of $Ro$.  Therefore, the value of $Ro$ fully determines the basic state of the flow, which then scales with $Re$ in the same manner as the case of outer-stationary Taylor-Couette flow ($Ro = \infty$).

The behavior of $G/G_{\infty}$ is distinct in the four regions of the parameter space.  For Rayleigh-stable flows (regions I and II), $G/\Gi$ is less than 0.22.  On the other hand, for $-0.37 < Ro^{-1} < 0$ the torque is enhanced.  Since $G/G_{\infty}$ scales as $Ro^{-1}$ within each region,  we perform linear regressions of $G/G_{\infty}$ as a function of $Ro^{-1}$.  The resulting functions are shown as solid lines in Fig.\ \ref{Gfig}(b):
\begin{eqnarray}
\label{Gfit1}
\mr{region~I:~~} G/\Gi = 0.078 Ro^{-1} + 0.27~\\
\mr{region~II:~\:} G/\Gi = 0.17 \pm 0.05~~~~~~~~~\,\\ \label{Gfit2}
\mr{region~III:~~} G/\Gi = -0.95 Ro^{-1} + 1.00\\ \label{Gfit3}
\mr{region~IV:~~} G/\Gi = 2.24 Ro^{-1} + 1.83~~ \label{Gfit4}
\end{eqnarray}

\begin{figure}
\begin{center}
\includegraphics[width=8.6cm]{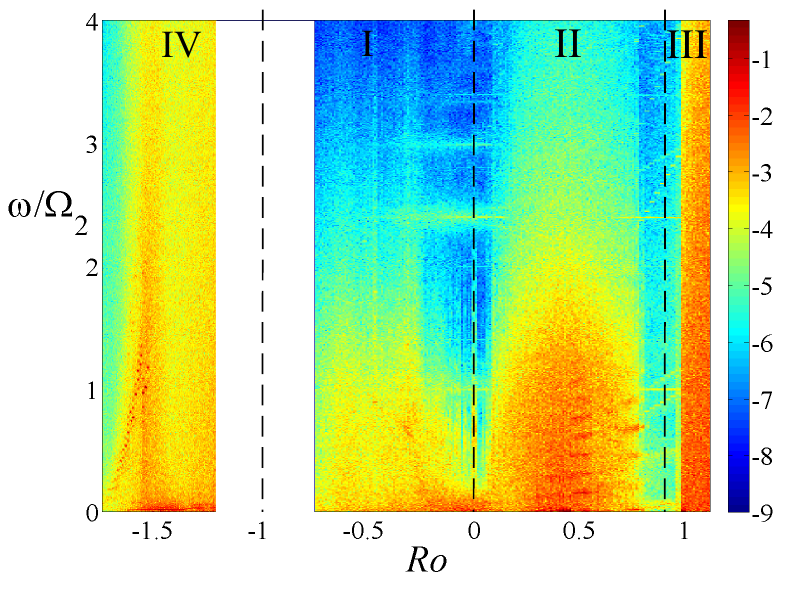}%
\caption{Fluctuations in the wall shear-stress strongly depend upon $Ro$.  The logarithmic spectral power density is indicated by color and the spectral frequency $\omega$ is normalized by $\Omega_2/2\pi = 8.22$~s\inv.  The experimentally inaccessible range of $-1.25 < Ro < 0.75$ is shown as white.}
\label{spectro}
\end{center}
\end{figure}

In addition to the distinct torque scaling in regions~I--IV, the wall shear-stress spectra also show marked changes with $Ro$.  Figure \ref{spectro} shows a spectrogram for $-2 < Ro < 2.1$.  For Rayleigh-stable flows $\(-1 < Ro < 0.905\)$, the system is characterized by narrow-band, weak shear stress fluctuations.  The fluctuations are stronger in region~II, which also shows strong wave modes near $Ro = 0.5$.  Region~III only has strong, broadband fluctuations for $Ro > 0.95$, even though the system becomes linearly unstable at $Ro = 0.905$ in the case of vanishing viscosity.  Finally, the spectra for flows in region~IV are also broadband but with much stronger wave modes evident.

Our measurements of the dimensionless torque $G$ and wall shear stress spectra indicate that the dominant control parameter for rotating shear flows is the Rossby number $Ro$.  We have not observed any transitions or evidence for nonlinear instabilities with increasing $Re$ in regions~I, III or IV, although we have indications of hysteresis in region~II.  This likely indicates that quasi-Keplerian flows can be nonlinearly unstable, but more systematic studies are needed to determine if this is indeed the case.

\begin{figure}
\begin{center}
\includegraphics[width=8.3cm]{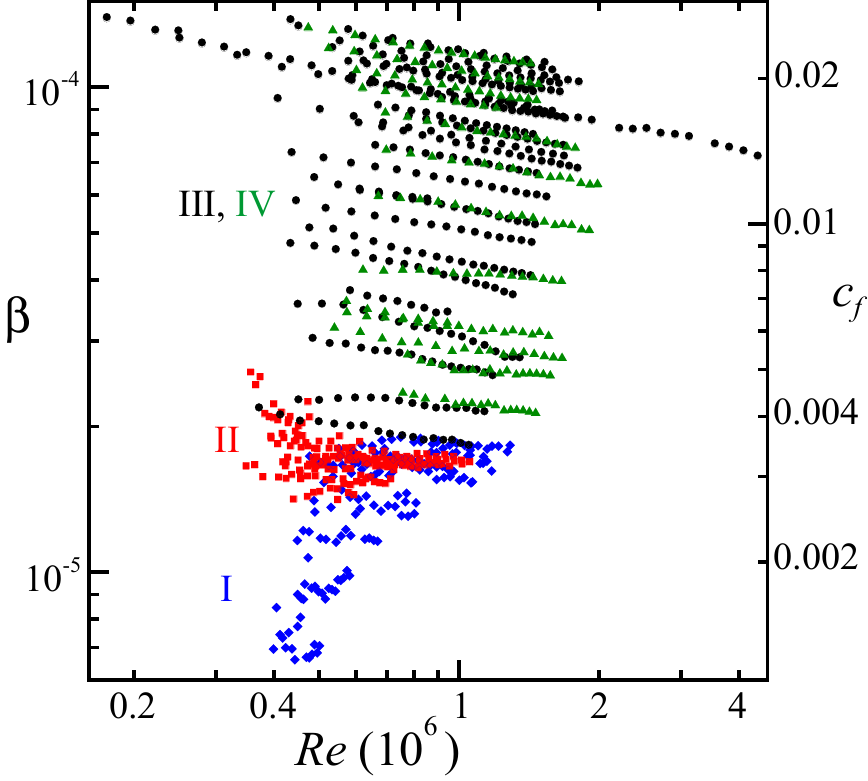}%
\caption{Scaling of $\beta$, given in eq.~(\ref{beta}), and the skin friction coefficient $c_f = G/Re^2$ with $Re$. The symbols correspond to the portions of parameter space defined in Fig.\ \ref{param_space}.}
\label{beta_fig}
\end{center}
\end{figure}

In ref.\ \cite{eckhardt07a} the variations of the torque with rotation rates are modeled using exact relations derived from the Navier-Stokes equations and assumptions about the torque contributions from the radial and vertical velocity fields (the \lq \lq wind").  Our observed variations with $Ro$ suggest that, while the exact relations in ref.\ \cite{eckhardt07a} contain only $\Omega_1 - \Omega_2$, there is a dependence of the wind on $Ro$.  The data presented here may be used to determine this dependence and provide testable predictions for the velocity fields (which are not presently accessible).

Ji \et \cite{ji06} did not directly measure $G$, but instead used their velocity measurements to determine the parameter $\beta$, which has been used to interpret angular momentum transport in astrophysical objects \cite{richard99,hure01,richard03,dubrelle05}.  In this prescription $\beta \equiv \overline{v_{\theta}\prime v_r}\prime/\(\overline{v_{\theta}}^2 q^2\)$, where overbars indicate mean quantities, $v_{\theta}$ is the azimuthal velocity, $v_{\theta}\prime = v_{\theta} - \overline{v_{\theta}}$, $v_r\prime = v_r - \overline{v_r}$ and $q = -\partial \ln \Omega / \partial \ln r$.  This model describes angular momentum transport as a diffusive process with a diffusivity $\nu_{\mr{turb}} = \beta \left|r^3 \partial \Omega /\partial r \right|$.  Thus, for a given $\Omega(r)$, $\beta$ determines the transport of angular momentum, with larger values of $\beta$ corresponding to larger fluxes of $L$.  For quasi-Keplerian flows (region~II), Ji \et measured $\beta = \(0.72 \pm 2.7\) \times 10^{-6}$.  This value is smaller than the value of $\beta = \(1.5 \pm 0.5\)\times 10^{-5}$ determined by Richard and Zahn \cite{richard99} for $Ro = -1$ ($\Omega_1 = 0$).  Ji \et attributed the disparity to Ekman circulation produced at the axial boundaries, which is reduced in their experiments by independently controlling the angular velocity of the axial boundary \cite{schartman09}.

To determine $\beta$ from our measurements of $G$ we use the expression given by Dubrulle \et \cite{dubrelle05}
\begin{equation}
\beta = \frac{1}{2\pi}R_C^4 \frac{G}{Re^2}\frac{S_{\mr{lam}}}{\overline{S}},
\label{beta}
\end{equation}
where $R_C = 2(b-a)/(b+a)$ and we take $S_{\mr{lam}}/\overline{S} = 3$ from \cite{dubrelle05}.  Our measured values of $\beta$ are shown in Fig.\ \ref{beta_fig}.  We note that $\beta$ is proportional to the skin friction coefficient $c_f = G/Re^2$, which is denoted on the right vertical axis.  The values of $\beta$ are shifted vertically for a given $Ro$ by the same factor as $G/\Gi$ shown in Fig.\ \ref{Gfig}(b).  For $Ro = -1$ we measure $\beta = (1.84 \pm 0.03) \times 10^{-5}$, which agrees with the value determined in ref.\ \cite{richard99}.  For the flows in region~II we determine an average value of $\overline{\beta} = (1.7 \pm 0.2) \times 10^{-5}$, which is markedly higher than the value given in ref.\ \cite{ji06}.  For Rayleigh-unstable flows, though, Ji \et report $\beta > 10^{-3}$ \cite{ji06} whereas our values span $2 \times 10^{-5} < \beta < 2 \times 10^{-4}$.

In conclusion, we have presented the first characterization of the flux of angular momentum (torque) between independently rotating cylinders for all regions of parameter space.  The reduction or enhancement of the torque $G/\Gi$ at a given Reynolds number only depends upon the Rossby number $Ro$.  The $Ro$-dependence of $G/\Gi$ is well described by eqns. (\ref{Gfit1})--(\ref{Gfit4}).  In contrast to ref.\ \cite{ji06} but in agreement with ref.\ \cite{richard99}, our measurements of $\beta$, which may be used to model angular momentum transport, are nonzero for Rayleigh-stable flows.  This disparity likely indicates that multiple states are possible for Rayleigh-stable flows, with our measurements representing a \lq \lq turbulent state" and those in ref.\ \cite{ji06} a \lq \lq laminar state."  This is particularly important for astrophysical flows where such nonlinear instability could explain the observed angular momentum transport.  Systematically perturbing Rayleigh-stable flows while measuring the torque could be used to directly test for nonlinear instabilities.

\begin{acknowledgments}
 We would like to thank B.~Eckhardt, Michael E.~Fisher, C.~Kalelkar, D.~Lohse, D.~Martin, H.~L.~Swinney and D.~S.~Zimmerman and the support of NSF-DMR 0906109.
 \end{acknowledgments}


\begin{thebibliography}{21}%
\makeatletter
\providecommand \@ifxundefined [1]{%
 \@ifx{#1\undefined}
}%
\providecommand \@ifnum [1]{%
 \ifnum #1\expandafter \@firstoftwo
 \else \expandafter \@secondoftwo
 \fi
}%
\providecommand \@ifx [1]{%
 \ifx #1\expandafter \@firstoftwo
 \else \expandafter \@secondoftwo
 \fi
}%
\providecommand \natexlab [1]{#1}%
\providecommand \enquote  [1]{``#1''}%
\providecommand \bibnamefont  [1]{#1}%
\providecommand \bibfnamefont [1]{#1}%
\providecommand \citenamefont [1]{#1}%
\providecommand \href@noop [0]{\@secondoftwo}%
\providecommand \href [0]{\begingroup \@sanitize@url \@href}%
\providecommand \@href[1]{\@@startlink{#1}\@@href}%
\providecommand \@@href[1]{\endgroup#1\@@endlink}%
\providecommand \@sanitize@url [0]{\catcode `\\12\catcode `\$12\catcode
  `\&12\catcode `\#12\catcode `\^12\catcode `\_12\catcode `\%12\relax}%
\providecommand \@@startlink[1]{}%
\providecommand \@@endlink[0]{}%
\providecommand \url  [0]{\begingroup\@sanitize@url \@url }%
\providecommand \@url [1]{\endgroup\@href {#1}{\urlprefix }}%
\providecommand \urlprefix  [0]{URL }%
\providecommand \Eprint [0]{\href }%
\@ifxundefined \urlstyle {%
  \providecommand \doi  [0]{\begingroup \@sanitize@url \@doi}%
  \providecommand \@doi [1]{\endgroup \@@startlink {\doibase
  #1}doi:\discretionary {}{}{}#1\@@endlink }%
}{%
  \providecommand \doi  [0]{doi:\discretionary{}{}{}\begingroup
  \urlstyle{rm}\Url }%
}%
\providecommand \doibase [0]{http://dx.doi.org/}%
\providecommand \Doi [0]{\begingroup \@sanitize@url \@Doi }%
\providecommand \@Doi  [1]{\endgroup\@@startlink{\doibase#1}\@@Doi}%
\providecommand \@@Doi [1]{#1\@@endlink}%
\providecommand \selectlanguage [0]{\@gobble}%
\providecommand \bibinfo  [0]{\@secondoftwo}%
\providecommand \bibfield  [0]{\@secondoftwo}%
\providecommand \translation [1]{[#1]}%
\providecommand \BibitemOpen [0]{}%
\providecommand \bibitemStop [0]{}%
\providecommand \bibitemNoStop [0]{.\EOS\space}%
\providecommand \EOS [0]{\spacefactor3000\relax}%
\providecommand \BibitemShut  [1]{\csname bibitem#1\endcsname}%
\bibitem [{\citenamefont {{Shakura}}\ and\ \citenamefont
  {{Sunyaev}}(1973)}]{shakura73}%
  \BibitemOpen
  \bibfield  {author} {\bibinfo {author} {\bibfnamefont {N.~I.}\ \bibnamefont
  {{Shakura}}}\ and\ \bibinfo {author} {\bibfnamefont {R.~A.}\ \bibnamefont
  {{Sunyaev}}},\ }\href@noop {} {\bibfield  {journal} {\bibinfo  {journal}
  {Astron. Astrophys.},\ }\textbf {\bibinfo {volume} {24}},\ \bibinfo {pages}
  {337} (\bibinfo {year} {1973})}\BibitemShut {NoStop}%
\bibitem [{\citenamefont {{Lathrop}}\ \emph
  {et~al.}(1992){\natexlab{a}}\citenamefont {{Lathrop}}, \citenamefont
  {{Fineberg}},\ and\ \citenamefont {{Swinney}}}]{lathrop92}%
  \BibitemOpen
  \bibfield  {author} {\bibinfo {author} {\bibfnamefont {(a) D.~P.}\ \bibnamefont
  {{Lathrop}}}, \bibinfo {author} {\bibfnamefont {J.}~\bibnamefont
  {{Fineberg}}}, \ and\ \bibinfo {author} {\bibfnamefont {H.~L.}\ \bibnamefont
  {{Swinney}}},\ }\Doi {10.1103/PhysRevLett.68.1515} {\bibfield  {journal}
  {\bibinfo  {journal} {Phys. Rev. Lett.},\ }\textbf {\bibinfo {volume} {68}},\
  \bibinfo {pages} {1515}}\BibitemShut
  {NoStop}; {\bibfield  {journal}
  {\bibinfo  {journal} {(b) \pra},\ }\textbf {\bibinfo {volume} {46}},\ \bibinfo
  {pages} {6390} (\bibinfo {year} {1992}{\natexlab{b}})}\BibitemShut {NoStop}%
\bibitem [{\citenamefont {{Lewis}}\ and\ \citenamefont
  {{Swinney}}(1999)}]{lewis99}%
  \BibitemOpen
  \bibfield  {author} {\bibinfo {author} {\bibfnamefont {G.~S.}\ \bibnamefont
  {{Lewis}}}\ and\ \bibinfo {author} {\bibfnamefont {H.~L.}\ \bibnamefont
  {{Swinney}}},\ }\Doi {10.1103/PhysRevE.59.5457} {\bibfield  {journal}
  {\bibinfo  {journal} {\pre},\ }\textbf {\bibinfo {volume} {59}},\ \bibinfo
  {pages} {5457} (\bibinfo {year} {1999})}\BibitemShut {NoStop}%
\bibitem [{\citenamefont {{Richard}}\ and\ \citenamefont
  {{Zahn}}(1999)}]{richard99}%
  \BibitemOpen
  \bibfield  {author} {\bibinfo {author} {\bibfnamefont {D.}~\bibnamefont
  {{Richard}}}\ and\ \bibinfo {author} {\bibfnamefont {J.}~\bibnamefont
  {{Zahn}}},\ }\href@noop {} {\bibfield  {journal} {\bibinfo  {journal}
  {Astron. Astrophys.},\ }\textbf {\bibinfo {volume} {347}},\ \bibinfo {pages}
  {734} (\bibinfo {year} {1999})}\BibitemShut {NoStop}%
\bibitem [{\citenamefont {{Hur{\'e}}}\ \emph {et~al.}(2001)\citenamefont
  {{Hur{\'e}}}, \citenamefont {{Richard}},\ and\ \citenamefont
  {{Zahn}}}]{hure01}%
  \BibitemOpen
  \bibfield  {author} {\bibinfo {author} {\bibfnamefont {J.}~\bibnamefont
  {{Hur{\'e}}}}, \bibinfo {author} {\bibfnamefont {D.}~\bibnamefont
  {{Richard}}}, \ and\ \bibinfo {author} {\bibfnamefont {J.}~\bibnamefont
  {{Zahn}}},\ }\Doi {10.1051/0004-6361:20000536} {\bibfield  {journal}
  {\bibinfo  {journal} {Astron. Astrophys.},\ }\textbf {\bibinfo {volume}
  {367}},\ \bibinfo {pages} {1087} (\bibinfo {year} {2001})}\BibitemShut
  {NoStop}%
\bibitem [{\citenamefont {{Richard}}(2003)}]{richard03}%
  \BibitemOpen
  \bibfield  {author} {\bibinfo {author} {\bibfnamefont {D.}~\bibnamefont
  {{Richard}}},\ }\Doi {10.1051/0004-6361:20031010} {\bibfield  {journal}
  {\bibinfo  {journal} {Astron. Astrophys.},\ }\textbf {\bibinfo {volume}
  {408}},\ \bibinfo {pages} {409} (\bibinfo {year} {2003})}\BibitemShut
  {NoStop}%
\bibitem [{\citenamefont {{Dubrulle \textit{et al.}}}(2005)}]{dubrelle05}%
  \BibitemOpen
  \bibfield  {author} {\bibinfo {author} {\bibfnamefont {B.}~\bibnamefont
  {{Dubrulle \textit{et al.}}}},\ }\Doi {10.1063/1.2008999} {\bibfield
  {journal} {\bibinfo  {journal} {Phys. Fluids},\ }\textbf {\bibinfo {volume}
  {17}},\ \bibinfo {pages} {095103} (\bibinfo {year} {2005})}\BibitemShut
  {NoStop}%
\bibitem [{\citenamefont {{Ji \textit{et al.}}}(2006)}]{ji06}%
  \BibitemOpen
  \bibfield  {author} {\bibinfo {author} {\bibnamefont {{Ji \textit{et
  al.}}}},\ }\Doi {10.1038/nature05323} {\bibfield  {journal} {\bibinfo
  {journal} {\nat},\ }\textbf {\bibinfo {volume} {444}},\ \bibinfo {pages}
  {343} (\bibinfo {year} {2006})}\BibitemShut {NoStop}%
\bibitem [{\citenamefont {{Eckhardt}}\ \emph {et~al.}(2007)\citenamefont
  {{Eckhardt}}, \citenamefont {{Grossmann}},\ and\ \citenamefont
  {{Lohse}}}]{eckhardt07a}%
  \BibitemOpen
  \bibfield  {author} {\bibinfo {author} {\bibfnamefont {B.}~\bibnamefont
  {{Eckhardt}}}, \bibinfo {author} {\bibfnamefont {S.}~\bibnamefont
  {{Grossmann}}}, \ and\ \bibinfo {author} {\bibfnamefont {D.}~\bibnamefont
  {{Lohse}}},\ }\Doi {10.1017/S0022112007005629} {\bibfield  {journal}
  {\bibinfo  {journal} {J. Fluid Mech.},\ }\textbf {\bibinfo {volume} {581}},\
  \bibinfo {pages} {221} (\bibinfo {year} {2007})}\BibitemShut {NoStop}%
\bibitem [{\citenamefont {{Ravelet}}\ \emph {et~al.}(2010)\citenamefont
  {{Ravelet}}, \citenamefont {{Delfos}},\ and\ \citenamefont
  {{Westerweel}}}]{ravelet10}%
  \BibitemOpen
  \bibfield  {author} {\bibinfo {author} {\bibfnamefont {F.}~\bibnamefont
  {{Ravelet}}}, \bibinfo {author} {\bibfnamefont {R.}~\bibnamefont {{Delfos}}},
  \ and\ \bibinfo {author} {\bibfnamefont {J.}~\bibnamefont {{Westerweel}}},\
  }\Doi {10.1063/1.3392773} {\bibfield  {journal} {\bibinfo  {journal} {Phys.
  Fluids},\ }\textbf {\bibinfo {volume} {22}},\ \bibinfo {pages} {055103}
  (\bibinfo {year} {2010})}\BibitemShut {NoStop}%
\bibitem [{\citenamefont {{Wendt}}(1933)}]{wendt33}%
  \BibitemOpen
  \bibfield  {author} {\bibinfo {author} {\bibfnamefont {F.}~\bibnamefont
  {{Wendt}}},\ }\href@noop {} {\bibfield  {journal} {\bibinfo  {journal} {Ing.
  Arch.},\ }\textbf {\bibinfo {volume} {4}},\ \bibinfo {pages} {577} (\bibinfo
  {year} {1933})}\BibitemShut {NoStop}%
\bibitem [{\citenamefont {{Taylor}}(1936)}]{taylor36}%
  \BibitemOpen
  \bibfield  {author} {\bibinfo {author} {\bibfnamefont {G.~I.}\ \bibnamefont
  {{Taylor}}},\ }\href@noop {} {\bibfield  {journal} {\bibinfo  {journal} {Roy.
  Soc. London Proc. A},\ }\textbf {\bibinfo {volume} {157}},\ \bibinfo {pages}
  {546} (\bibinfo {year} {1936})}\BibitemShut {NoStop}%
\bibitem [{\citenamefont {{Rayleigh}}(1916)}]{rayleigh16}%
  \BibitemOpen
  \bibfield  {author} {\bibinfo {author} {\bibfnamefont {L.}~\bibnamefont
  {{Rayleigh}}},\ }\href@noop {} {\bibfield  {journal} {\bibinfo  {journal}
  {Proc. Roy. Soc. Lond. A},\ }\textbf {\bibinfo {volume} {93}},\ \bibinfo
  {pages} {148} (\bibinfo {year} {1916})}\BibitemShut {NoStop}%
\bibitem [{\citenamefont {{Coles}}(1965)}]{coles65}%
  \BibitemOpen
  \bibfield  {author} {\bibinfo {author} {\bibfnamefont {D.}~\bibnamefont
  {{Coles}}},\ }\Doi {10.1017/S0022112065000241} {\bibfield  {journal}
  {\bibinfo  {journal} {J. Fluid Mech.},\ }\textbf {\bibinfo {volume} {21}},\
  \bibinfo {pages} {385} (\bibinfo {year} {1965})}\BibitemShut {NoStop}%
\bibitem [{\citenamefont {{Velikhov}}(1959)}]{velikhov59}%
  \BibitemOpen
  \bibfield  {author} {\bibinfo {author} {\bibfnamefont {E.~P.}\ \bibnamefont
  {{Velikhov}}},\ }\href@noop {} {\bibfield  {journal} {\bibinfo  {journal}
  {Sov. Phys. JETP},\ }\textbf {\bibinfo {volume} {9}},\ \bibinfo {pages} {995}
  (\bibinfo {year} {1959})}\BibitemShut {NoStop}%
\bibitem [{\citenamefont {{Chandrasekhar}}(1960)}]{chandrasekhar60}%
  \BibitemOpen
  \bibfield  {author} {\bibinfo {author} {\bibfnamefont {S.}~\bibnamefont
  {{Chandrasekhar}}},\ }\Doi {10.1073/pnas.46.2.253} {\bibfield  {journal}
  {\bibinfo  {journal} {Proc. Natl. Acad. Sci. U.S.A},\ }\textbf {\bibinfo
  {volume} {46}},\ \bibinfo {pages} {253} (\bibinfo {year} {1960})}\BibitemShut
  {NoStop}%
\bibitem [{\citenamefont {{Balbus}}\ and\ \citenamefont
  {{Hawley}}(1991)}]{balbus91}%
  \BibitemOpen
  \bibfield  {author} {\bibinfo {author} {\bibfnamefont {(a) S.~A.}\ \bibnamefont
  {{Balbus}}}\ and\ \bibinfo {author} {\bibfnamefont {J.~F.}\ \bibnamefont
  {{Hawley}}},\ }\Doi {10.1086/170270} {\bibfield  {journal} {\bibinfo
  {journal} {\apj},\ }\textbf {\bibinfo {volume} {376}},\ \bibinfo {pages}
  {214} (\bibinfo {year} {1991})}\BibitemShut {NoStop}; %
  \Doi {10.1103/RevModPhys.70.1} {\bibfield  {journal} {\bibinfo
   {journal} {(b) Rev. Mod. Phys.},\ }\textbf {\bibinfo {volume} {70}},\ \bibinfo
  {pages} {1} (\bibinfo {year} {1998})}\BibitemShut {NoStop}%
\bibitem [{\citenamefont {{Sisan \textit{et al.}}}(2004)}]{sisan04}%
  \BibitemOpen
  \bibfield  {author} {\bibinfo {author} {\bibfnamefont {D.~R.}\ \bibnamefont
  {{Sisan \textit{et al.}}}},\ }\Doi {10.1103/PhysRevLett.93.114502} {\bibfield
   {journal} {\bibinfo  {journal} {Phys. Rev. Lett.},\ }\textbf {\bibinfo
  {volume} {93}},\ \bibinfo {pages} {114502} (\bibinfo {year}
  {2004})}\BibitemShut {NoStop}%
\bibitem [{\citenamefont {{Schartman}}\ \emph {et~al.}(2009)\citenamefont
  {{Schartman}}, \citenamefont {{Ji}},\ and\ \citenamefont
  {{Burin}}}]{schartman09}%
  \BibitemOpen
  \bibfield  {author} {\bibinfo {author} {\bibfnamefont {E.}~\bibnamefont
  {{Schartman}}}, \bibinfo {author} {\bibfnamefont {H.}~\bibnamefont {{Ji}}}, \
  and\ \bibinfo {author} {\bibfnamefont {M.~J.}\ \bibnamefont {{Burin}}},\
  }\Doi {10.1063/1.3077942} {\bibfield  {journal} {\bibinfo  {journal} {Rev.
  Sci. Instrum.},\ }\textbf {\bibinfo {volume} {80}},\ \bibinfo {pages}
  {024501} (\bibinfo {year} {2009})}\BibitemShut {NoStop}%
\end{thebibliography}

%

\end{document}